\documentstyle[12pt]{article}
\newcommand{\clequ}{\setcounter{equation}{0}}
\begin{document}

\begin{titlepage}
\begin{center}
\ \\
\vskip 3.cm
{\Large\bf Reduction of three-band model for copper oxides to single-band
generalized t~-~J model }
\vskip 2.cm
{\large  V.I.Belinicher and A.L.Chernyshev } \\

Institute of Semiconductor Physics, 630090 Novosibirsk, Russia\\
\end{center}
\vskip 2.cm

\begin{abstract}
A three-band model for copper oxides in the region of parameters
where the second hole on the copper has energy close to the first hole
on the oxygen is considered. The exact solution for one hole on a
ferromagnetic background of the ordered copper spins is obtained. A general
procedure for transformation of the primary Hamiltonian to the Hamiltonian
of singlet and triplet excitations is proposed. Reduction of the
singlet-triplet Hamiltonian to the single-band Hamiltonian of the
generalized t~-~J model is performed. A comparison of the solution
for the generalized t~-~J model on a ferromagnetic background with
the exact solution shows a very good agreement.
\end{abstract}

\end{titlepage}

\section{ Introduction}
\clequ \

Some time ago the extended Hubbard model or the Emery model was
proposed for description of holes in \(CuO_{2}\) plane \cite{Em1}.
The next essential step was made by Zhang and Rice \cite{Zh1}.
They proposed that holes on the oxygen move over the crystal in the form
of spin-singlets formed with the copper spins and can be described by
the single-band t~-~J model. It should be noted that consistent
reduction of the three-band model Hamiltonian to the
single-band model t~-~J Hamiltonian was not presented in \cite{Zh1}.
Therefore polemic concerning validity of the t~-~J model
has arisen.

In the work by Emery and Reiter \cite{Em2}, the exact
solution of the three-band model on a ferromagnetic background of
Cu-spins was obtained. They have shown that this exact solution
can be interpreted in the region of small momenta as motion
of a spin triplet formed by the O-hole and two adjacent
Cu-spins. Zhang and Rice \cite{Zh2} have shown that the exact
solution on a ferromagnetic background can be interpreted as
motion of the local spin-singlet. In the work by Zhang \cite{Zh3}
it has been shown that the spectra of the t~-~J model and of the three-band
model are identical: if the eigenstate of the t~-~J model is
known one can construct the eigenstate of the three-band model
with the same energy with the help of the local spin-singlet.
However this does not mean the physical equivalence of the two models
because the wave functions of the local spin-singlets are not
orthogonal, as it was stressed by Emery and Reiter \cite{Em3}.

The effective Hamiltonian  in  terms  of  singlet  and  triplet
operators was obtained by Shen and Ting \cite{Sh1}. The  contribution  of
the triplet state was estimated to be of the order  of  10\% on  an
antiferromagnetic background. This value determines the  precision
 of the single-band approximation.  Notice that  it is sufficiently
difficult to recognize the  Hamiltonian of the  t~-~J model
in the final formula of \cite{Sh1}.  The work by Pang,
  Xiang,  Su and  Lu \cite{Pa1} was
devoted to the construction  of  the  singlet  and  triplet  states  and to a
comparison of the hopping parameters on a ferromagnetic background with
the exact solution \cite{Em2}. A sufficiently good agreement was obtained.

All above-named works  were  dealing with parameters  of the Emery
model  \cite{Em1} in the region $U_{d}-\epsilon,
\epsilon \gg |t|$, where $U_{d}$ is the Coulomb repulsion at the Cu site,
$\epsilon$ is the difference in  energy  between the O(2p) and Cu(3d) holes
and t is the Cu - O hopping parameter.  This condition means that the
energy of the $p_x,p_y$ oxygen  levels
lies between and sufficiently  far from the energy of the $d_{x^{2}-y^{2}}$
copper  levels splitted by  the Coulomb  repulsion  $U_{d}$.

A  more  accurate
estimation (see work by Lovtsov and Yushankhai \cite{Yu1} and  this
work below) shows that in  fact  the  condition  of  applicability of  the
perturbation theory is more rigid: $U_{d}-\epsilon,\epsilon \gg 4\sqrt{2}|t|$.
 Different  band  calculations  \cite{Pi1,Fl1}  give $t \approx$ -1.4eV
and the perturbation theory over $|t|$  for  computing the
properties  of  charge
carriers works at $U_{d}> 16$eV. Known estimations \cite{Pi1,Fl1}
give $U_{d} \leq$ 8eV.
The situation is more simple if oxygen levels are  close  to  the
lower or the higher $d_{x^{2}-y^{2}}$ copper level . We use the hole
classification of the energy levels.  The case $\epsilon \ll U_{d}$ was
considered in the work by
Lovtsov and Yushankhai \cite{Yu1},  where local singlet and triplet  states
were constructed and  hopping  of  these  states  over the crystal  was
studied.

     In this work we study the case $U_{d}-\epsilon \ll U_{d}$
when  the  oxygen level is close to the higher copper $d_{x^{2}-y^{2}}$
level.  Such  level
position was proposed as a result of band  calculations  in  the work by
Flambaum and Sushkov \cite{Fl1}  and does not contradict the photoemission
data \cite{Fi1}.   Actually,  the  difference  in  position  between the lower
$d_{x^{2}-y^{2}}$ and the $p_x,p_y$  levels is approximately 4eV\cite{Fi1} .
For $U_{d}=$8eV the $p_x,p_y$ levels are in the middle between the  splitted
Cu $d_{x^{2}-y^{2}}$ levels,
but for $U_{d}=6$eV  as proposed in \cite{Fl1}, the  O $p_x,p_y$ levels
are closer to
the higher $d_{x^{2}-y^{2}}$ level. In this  work  the  direct
oxygen-oxygen hopping is not taken into account.

 The work can be divided  into
three parts. In the first part we will get the exact solution
of the three-band model on a ferromagnetic  background  and  discuss
its properties.  This solution is  an analog of  the  corresponding
solution  of Emery and Reiter \cite{Em2}.   The consideration of the hole
motion on a ferromagnetic background is of certain methodical interest.
 This solution is exact but it describes not the ground state but the
high-excitation state. Such solution is used for testing the approximate
Hamiltonian of the generalized t-J model obtained in the present work
from the three-band Hamiltonian. This allows to make a simple estimation
of the magnitude of corrections to the t-J model which appear in the
reduction of the three-band model. An estimation of such corrections
 for the hole motion on an antiferromagnetic background of the copper spins
represents a separate problem.

In  the  second  part   of   the   work   we   transform   the
three-band Hamiltonian to  the  Hamiltonian  describing  hopping  and
transition between two singlet  and  one   triplet   states.
These singlet and triplet states are formed by the spin of the hole
and the spin of the copper.
  For performing the transformation the technique  of  representation  of
the Hubbard  operators in terms of the hole and the spin-$\frac{1}{2}$
  operators was used.
This Hamiltonian, also containing three bands (two singlet and one triplet),
with the help of the Schrieffer - Wolff transformation is reduced to the
low-energy Hamiltonian for the lower singlet.  It is  the  Hamiltonian
of  the generalized t~-~J model.

In the  third part of the work a detailed comparison of
the  properties  of the  generalized   t~-~J  model  on  a  ferromagnetic
background with the exact solution of  the  three-band  model  on  a
ferromagnetic  background  is  made.
Corrections to the t-J model Hamiltonian providing an agreement with the
exact result are estimated.
   An  excellent  agreement
between the approximate and exact solutions is shown.

In Appendix corrections to a single-band Hamiltonian  of  third-  and
fourth-order over nondiagonal hopping terms in the case $U_{d}=\epsilon$
are derived.

\section{ Exact solution for the three-band model on a
ferromagnetic background}
\clequ
\subsection{Three-band Hamiltonian and the exact solution} \

We want to remind that the Hamiltonian of the t~-~J model is usually
represented in the form
\begin{eqnarray}
\label{1tj}
&&H_{t-J}=t\sum_{<ll'>,\alpha}\tilde{c}^{+}_{l\alpha} \tilde{c}_{l'\alpha}+
J\sum_{ll'}{\bf S}_{l}{\bf S}_{l'},
\nonumber\\
&&\tilde{c}_{l\alpha} =c_{l\alpha}(1-\hat{n}_{l,-\alpha}), \ \
\tilde{c}^{+}_{l\alpha}=(\tilde{c}_{l\alpha})^{+}, \ \
\nonumber\\
&&{\bf S}_{l}=(1/2)c_{l}^{+}{\bf \sigma}c_{l}, \ \ \
\hat{n}_{l\alpha }=c_{l\alpha }^{+}c_{l\alpha }.
\end{eqnarray}
Here $c_{l\alpha}^{+},c_{l\alpha}$  are the electron creation and annihilation
operators at the lattice site $l$, $ \alpha = \uparrow, \downarrow $ or
$\pm \frac{1}{2}$ is the spin progection,
 ${\bf \sigma}$ are Pauly matrices, the symbol $<ll'>$ denotes
summation over the nearest-neighbors, t is the hopping integral, J is
the superexchange energy. It will be more convenient for us to use another
form of representation of the Hamiltonian (\ref{1tj}) in terms of Hubbard
operators.
\begin{equation}
\label{1}
H_{t~-~J}=E_{0}\sum_{l}X^{00}_{l}+t\sum_{<ll'>}X^{\alpha0}
_{l}X^{0\alpha}_{l'}+J\sum_{<ll'>}{\bf S}_{l}{\bf S}_{l'}
\end{equation}
here $X^{ab}_{l}$ are Hubbard operators at the site $l: \  X^{ab}_{l}=
|al><lb|$ for the states $|a>, |b> =|0>,|\alpha>$. The connection between
the Hamiltonians (\ref{1tj}), (\ref{1}) is given by the following relations:
\begin{equation}
\label{tjt}
X^{\alpha0}_{l} \Rightarrow \tilde{c}^{+}_{l\alpha}, \ \
X^{0\alpha}_{l} \Rightarrow \tilde{c}_{l\alpha}, \ \
{\bf S}_{l}=(1/2){\bf \sigma}_{\alpha\beta}X^{\alpha \beta}_{l}
 \Rightarrow (1/2)c_{l}^{+}{\bf \sigma}c_{l}.
\end{equation}
We have added in Eq.(\ref{1}) the first term which describes the energy
of the quenched hole.

In the case of half-filling, the Hamiltonian (\ref{1}) reduces to the
Heisenberg Hamiltonian, and for $J>0$ the antiferromagnetic state
is its ground state. However, ferromagnetic state is an eigenstate of
this Hamiltonian and we can easily get a simple exact eigenstate
$|{\bf k}>$ for $H_{tJ}$ with one hole over ferromagnetic
background
\begin{eqnarray}
\label{2}
|{\bf k}> =\sum_{l}\exp(i{\bf k}{\bf r}_{l})X^{0\downarrow}_{l}
|f>,\ \ \ \ \ \ |f>=\prod_{l}c^+_{l\downarrow}|0>
\nonumber\\
\epsilon_{\bf k}=E_{0}+4t\gamma _{\bf k}, \ \ \ \ \
\gamma _{\bf k}=(1/2)(cos(k_xa)+cos(k_ya))
\end{eqnarray}
where $|f>$ is the ferromagnetic state at half filling and all electron
spins down, $\epsilon_{\bf k}$ is the electron energy.
We will construct the exact solution for the state with one hole over a
ferromagnetic background for the three band model in the region of
parameters discussed above.

The Hamiltonian has the form:
\begin{eqnarray}
\label{3}
H=\epsilon^{0}_{d}\sum_{l,\alpha}n^{d}_{l\alpha}+
\epsilon^{0}_{p}\sum_{m,\alpha}n^{p}_{m\alpha}+
U_{d}\sum_{l}n^{d}_{l\uparrow}n^{d}_{l\downarrow}+
\nonumber\\
V\sum_{<lm>\alpha\beta}n^{d}_{l\alpha}n^{p}_{m\beta}+
t\sum_{<lm>\alpha}(d^{+}_{l\alpha}p_{m\alpha}+p^{+}_{m\alpha}d_{l\alpha}),
\end{eqnarray}
where $d^{+}_{l\alpha} (d_{l\alpha})$ creates (annihilates) the
$d_{x^{2}-y^{2}}$ hole of spin projection $\alpha$ at the Cu site l
and $p^{+}_{m\beta} (p_{m\beta})$ creates (annihilates) the $p_{\beta}$
hole of spin projection $\beta$ at the O site $m, n^{d}_{l\alpha}=
d^{+}_{l\alpha}d_{l\alpha}, n^{p}_{m\beta}=p^{+}_{m\beta}p_{m\beta}$.
The sign convention in the last term of Eqs. (\ref{3}) corresponds to
the change of the signs of wave functions in all the odd cells, which
corresponds to the quasimomentum redefinition $(k_x,k_y) \rightarrow
(k_{x}+\pi/a, k_{y}+\pi/a)$.

In the case of one hole over unit filling of the $d_{x^2-y^2}$ copper
states at each site, one can get the reduced Hamiltonian. Using the
representation of $d^{+}_{l\alpha}, d_{l\alpha}$ in terms of the Hubbard
operators $X^{\alpha0}_{l}, X^{0\alpha}_{l}, X^{\alpha2}_{l},
X^{2\alpha}_{l}$
\begin{eqnarray}
\label{X3}
d^{+}_{l\alpha}=X^{\alpha 0}_{l}+2\alpha X^{2-\alpha}_{l},\ \ \
d_{l\alpha}    =X^{0\alpha}_{l}+2\alpha X^{-\alpha2}_{l}
\end{eqnarray}
and omitting the contribution of the $X^{\alpha 0}_{l},
X^{0\alpha}_{l}$ operators one can get
\begin{eqnarray}
\label{4}
&&H_{pd}=\epsilon_{p}\sum_{m\alpha}n^{p}_{m\alpha}+
\epsilon_{d}\sum_{l}X^{22}_{l}+
\nonumber\\
&&t\sum_{<lm>\alpha}2\alpha(p^{+}_{m\alpha}X^{-\alpha2}_{l}+
X^{2-\alpha}_{l}p_{m\alpha})
\end{eqnarray}
where $\epsilon_{d}=\epsilon^{0}_{d}+U_{d}$ and
$\epsilon_{p}=\epsilon^{0}_{p}+2V$ are the renormalized energies
of d and p states,\\ $X^{22}_{l}=|2l><l2|, X^{\alpha 2}_{l}=
|1\alpha l><l2|, X^{2\alpha}_{l}=|2 l><l \alpha|$ are the Hubbard
operators for the $d_{x^2-y^2}$ Cu states at the site $l$.
The Hubbard operators
$X^{22}_{l}, X^{\alpha 2}_{l}, X^{2\alpha}_{l}$ can be expressed
in terms of the d-operators of the copper
\begin{eqnarray}
\label{5}
X^{22}_{l}=n^{d}_{l\uparrow}n^{d}_{l\downarrow},\ \
X^{\alpha 2}_{l}=d^{+}_{l\alpha}d_{l\downarrow}d_{l\uparrow},\ \
X^{2\alpha}_{l}=d^{+}_{l\uparrow}d^{+}_{l\downarrow}d_{l\alpha}.
\end{eqnarray}
The Hamiltonian in the form similar to (\ref{3}) was used in many
works where the slave boson (fermion) method was applied to the three band
model \cite{Ca1,Gr1,Gr2,Fe1}. We will show that eigenstates of the
Hamiltonian (\ref{3}) on a ferromagnetic background can be represented in
the form
\begin{eqnarray}
\label{6}
&&|{\bf k}> =\sum_{l}\exp(i{\bf k}{\bf r}_{l})\hat{Y}_{l}({\bf k})|f>,
\ \ \ \ |f>=\prod_{l}d^{+}_{l\downarrow}|0>,
\nonumber\\
&&\hat{Y}_{l}({\bf k})=(\alpha({\bf k})d^{+}_{l\uparrow}d^{+}_{l\downarrow}
+\beta({\bf k})\pi_{l})d_{l\downarrow},\ \ \ \
|pl>=\pi_{l}d_{l\downarrow}|f>,
\nonumber\\
&&\pi_{l}=(1/\sqrt{2})(P^{+}_{l\uparrow}d^{+}_{l\downarrow}-
P^{+}_{l\downarrow}d^{+}_{l\uparrow}),\ \ \
P_{l\alpha}=(1/2)\sum_{m\in<l>}p_{m\alpha},
\end{eqnarray}
where $<l>$ are the nearest-neighbor sites to the site $l, |0>$ is the
vacuum state of the $CuO_{2}$ plane that corresponds to the completely filled
$3d^{10}$ shell of Cu and the $2p^{6}$ shell of O. The energy of this
state $|{\bf k}>$ is equal to
\begin{eqnarray}
\label{7}
&&\epsilon({\bf k})=\bar{\epsilon}-R({\bf k}),\ \ \
R({\bf k})=\sqrt{\Delta^{2}+8t^2\eta^{2}({\bf k})},
\nonumber\\
&&\bar{\epsilon}=(\epsilon_{d}+\epsilon_{p})/2,\ \ \
\Delta=(\epsilon_{d}-\epsilon_{p})/2,\ \ \
\nonumber\\
&&\eta({\bf k})^{2}=1+(1/2)\gamma_{{\bf k}},\ \ \
\gamma_{{\bf k}}=(1/2)(cos(k_{x}a)+cos(k_{y}a))
\end{eqnarray}
where $a$ is the distance between the Cu sites. Below we will count the energy
from $\bar{\epsilon}$. The coefficients $\alpha({\bf k}),\beta({\bf k})$
have the form
\begin{eqnarray}
\label{8}
\alpha({\bf k})=-\sqrt{(R({\bf k})-\Delta)/2R({\bf k})},\ \ \
\beta({\bf k})=\eta({\bf k})^{-1}\sqrt{(R({\bf k})+
\Delta)/2R({\bf k})}
\end{eqnarray}
and satisfy the normalization condition
\begin{eqnarray}
\label{9}
|\alpha({\bf k})|^{2}+\eta({\bf k})^{2}|\beta({\bf k})|^{2}=1.
\end{eqnarray}
The normalization condition (\ref{9}) has a nontrivial form due to the fact
that the states $|pl>$ are not orthogonal
\begin{eqnarray}
\label{91}
<pl'|pl>=\delta_{ll'}+(1/8)\delta_{<ll'>},
\end{eqnarray}
where $\delta_{<ll'>}=1$ if $l$ and $l'$ are the nearest-neighbors and
vanish otherwise. One can explicitly prove that the state (\ref{6}) is an
eigenstate of the Hamiltonian (\ref{4}) . Acting by the
Hamiltonian (\ref{4}) on the state $|{\bf k}>$ given by (\ref{6})
one can get the expression (\ref{8}) for the coefficients
$\alpha({\bf k}),\beta({\bf k})$.

\subsection{Interpretation of the exact solution} \

This exact solution can be interpreted as the Bloch
wave formed by the linear combination of two local singlets.
One local singlet represents two holes on Cu. Other singlet consists of
one hole on Cu and another hole on O
(or more accurately of coherent
sum of the hole states on the oxygens nearest to the copper).
The structure of the $CuO_{2}$ plane is shown in Fig.1.

\vskip 0.3cm
\begin{center}
\setlength{\unitlength}{1mm}
\begin{picture}(84,44)
\multiput(0,0)(40,0){3}{\circle{3}}
\multiput(0,40)(40,0){3}{\circle{3}}
\multiput(20,20)(40,0){2}{\circle{3}}
\multiput(-2,18)(40,0){3}{\line(1,1){4}}
\multiput(-2,22)(40,0){3}{\line(1,-1){4}}
\multiput(40,0)(-20,20){2}{\line(1,1){20}}
\multiput(20,20)(20,20){2}{\line(1,-1){20}}
\end{picture}
\end{center}
\vskip 0.3cm
\noindent
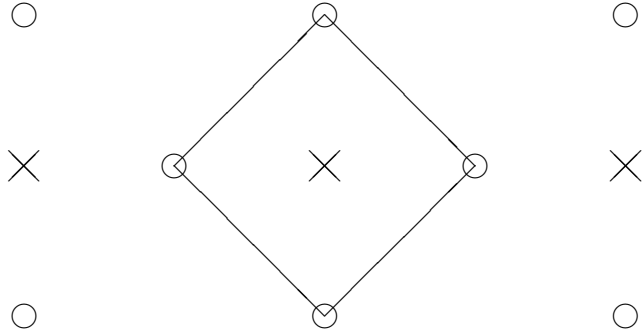
FIG.1 {\small The structure of the $CuO_{2}$ plane.
The crosses denote coppers,
circles denote oxygens. The local cluster is separated by solid lines.
The hole on the oxygens on the solid lines constitutes coherent state which
forms the local singlet.}
\vskip 0.3cm

Notice that at ${\bf k} \rightarrow 0$ the solution (\ref{7}) can be
represented in the form of Emery and Reiter triplets \cite{Em2}
\begin{eqnarray}
\label{10}
|{\bf k}>=\sum_{m,j=\pm 1}(\gamma d^{+}_{mj\uparrow}+\delta
(p^{+}_{mj\uparrow}-p^{+}_{m\downarrow}d^{+}_{mj\uparrow}
d_{mj\downarrow}))|f>\ \ \ \ \     at \ \ \ {\bf k} \rightarrow 0,
\end{eqnarray}
where summation over $m$ is produced over all oxygens in the $CuO_{2}$
plane and over $j$ on two adjacent to the oxygen coppers for
 $j=\pm1$. However, solution
(\ref{6}) can not be represented in the form (\ref{10}) for all
${\bf k}$.

The solution represents the sum of the overlapping singlets. Due to this
overlapping a spin density matrix $\rho_{O}$ of an oxygen hole has
a nontrivial form
\begin{eqnarray}
\label{11}
\rho_{O}=(1/2)(1+(\gamma_{{\bf k}}/(2+\gamma_{{\bf k}}))\sigma_{z}),
\end{eqnarray}
where $\gamma_{{\bf k}}$ is determined by Eq.(\ref{7}) and
$\sigma_z$ is the Pauly matrix. An average of oxygen hole spin
obtained with the help of the density matrix $\rho_{O}$ is equal
to 1/6 at ${\bf k} \rightarrow 0$.
This result corresponds to the calculations of Refs. \cite{Em2,Fl1}.

In the region of small ${\bf k}$ we have the following expression for
the energy of a singlet polaron
\begin{eqnarray}
\label{12}
\epsilon_{{\bf k}}=-R+(t^{2}/2R)k^{2}a^{2}+...,\ \ \
R=\sqrt{\Delta^{2}+12t^{2}}.
\end{eqnarray}
The gain in of energy in (\ref{12}) is sufficiently high: 12 is a
large number! According to the estimations of the works
\cite{Pi1,Fl1,Fi1} $\Delta \simeq 1 \div 2$eV, $t \simeq -1.4$eV and the
perturbation theory over $t/\Delta$ does not work:
 $\Delta^{2} \leq 4($eV$)^{2}$,
$\ 12t^{2} \simeq 24($eV$)^{2}\  $and$\  12t^{2} \gg \Delta^{2}$ and
we have at the small ${\bf k}$
\begin{eqnarray}
\label{13}
\epsilon_{{\bf k}}=-2\sqrt{3}t+t_{eff}k^{2}a^{2},\ \ \
t_{eff}=t/4\sqrt{3}=0.1443t \simeq 0.2eV.
\end{eqnarray}
The band width $w$ is equal to
\begin{eqnarray}
\label{14}
w=R-\sqrt{\Delta^{2}+4t^{2}} \simeq 2(\sqrt{3}-1)t = 1.469t,
\end{eqnarray}
which is 1.25 times the naive band width $2zt_{eff}=2t/\sqrt{3}$ . These
conclusions agree
with results of Flambaum and Sushkov obtained by variational method
\cite{Fl1}.
\subsection{Reformulation of the exact solution} \

In this section we reformulate the exact solution in other notations
which will be useful further. For this we produce a map of three
Cu states at every site $|1\downarrow>, |1\uparrow>, |2>=
|\uparrow\downarrow>$ into eight spin-hole states:
$|0,\alpha), |1\beta,\alpha), |2,\alpha)$ where $\alpha, \beta =\pm1/2$ or
$\downarrow, \uparrow$ are the spin-$\frac{1}{2}$  projections and the first
index is the number of holes. One can introduce the Fermi operators
for holes $h^{+}_{\alpha}, h_{\alpha}$ and the spin-$\frac{1}{2}$
operators ${\bf s}$, then
\begin{eqnarray}
\label{15}
s_{z}|0\alpha)=\alpha|0\alpha),\ \ \ |1\beta,\alpha)=h^{+}_{\beta}
|0\alpha),\ \ \ |2\alpha)=h^{+}_{\uparrow}
h^{+}_{\downarrow}|0\alpha).
\end{eqnarray}
The map has the following form:
\begin{eqnarray}
\label{16}
|1\alpha> \Rightarrow |0\alpha),\ \ \
|2> \Rightarrow  |s) \equiv (1/\sqrt{2})(|1\uparrow,\downarrow)-
|1\downarrow,\uparrow))
\end{eqnarray}
where $|s)$ is the hole-spin singlet state. This map generates the following
representation for Hubbard operators on copper (\ref{5}) \cite{Be1}
in terms of $h^{+}_{\alpha}, h_{\alpha}, {\bf s}$
\begin{eqnarray}
\label{17}
&&X^{2\alpha}=\sqrt{2}2\alpha (h^{+}\hat{s})_{-\alpha}(1-\hat{n}^{h}),
\ \ \ \hat{s}=(1-2{\bf s}{\bf \sigma})/4,
\nonumber\\
&&X^{\alpha 2}=\sqrt{2}2\alpha(1-\hat{n}^{h})(\hat{s}h)_{-\alpha},
\ \ \ \hat{n}^{h}=(h^{+}\cdot h),\ \  \
\hat{d}^{h}=h^{+}_{\uparrow}h_{\uparrow}h^{+}_{\downarrow}h_{\downarrow},
\nonumber\\
&&X^{22}=(h^{+}(1-\hat{n}^{h})\hat{s}h),
\ \ \ \hat{N}\equiv X^{\uparrow\uparrow}+X^{\downarrow\downarrow}=
1-\hat{n}^{h}+\hat{d}^{h}.
\end{eqnarray}

This representation can be used for description of one-hole states
if we omit the multiplier $(1-\hat{n}^{h})$ in Eq.(\ref{17}) for
$X^{2\alpha},X^{\alpha 2}$ and $X^{22}$ that is essential for replacing
the two-hole states $|2\alpha)$ (\ref{15}). Substituting the
representation (\ref{17})  for $X^{2\alpha},X^{\alpha 2}$ and $X^{22}$
in the Hamiltonian (\ref{4}) we can obtain a Hamiltonian in more
usual terms
\begin{eqnarray}
\label{18}
&&H_{pd}=\epsilon_{p}\sum_{m,\alpha}n^{p}_{m\alpha}+
\epsilon_{d}\sum_{l}h^{+}_{l}\hat{s}_{l}h_{l}+
\nonumber\\
&&\sqrt{2}t\sum_{l}(h^{+}_{l}\hat{s}_{l}P_{l}+P^{+}_{l}\hat{s}_{l}h_{l}).
\end{eqnarray}
This Hamiltonian contains the operators of holes
$p^{+}_{m\alpha}, p_{m\alpha}$ at the O sites, the operators of holes
$h^{+}_{\alpha}, h_{\alpha}$ at the Cu sites and the operators of
spin-$\frac{1}{2}$  ${\bf s}$ at the Cu sites. These operators act
on the ground state
where there is a spin-$\frac{1}{2}$ at every $Cu$ site. The operator
$\hat{s}_{l}$ represents a projector on the singlet state in one-particle
sector. The eigenstate of the Hamiltonian (\ref{4}), (\ref{18}) can
be represented as the sum of two singlets
\begin{eqnarray}
\label{19}
&&|{\bf k}> =\sum_{l}\exp(i{\bf k}{\bf r}_{l})\hat{Z}_{l}({\bf k})|f>,
\ \ \ \hat{Z}_{l}({\bf k})=\alpha({\bf k})\hat{S}^{d}_{l}+
\beta({\bf k})\hat{S}^{p}_{l},
\nonumber\\
&&\hat{S}^{d}_{l}=(1/\sqrt{2})(h^{+}_{l\uparrow}-
h^{+}_{l\downarrow}s^{+}_{l}),\ \ \
\hat{S}^{p}_{l}=(1/\sqrt{2})(P^{+}_{l\uparrow}-
P^{+}_{l\downarrow}s^{+}_{l}),
\end{eqnarray}
where $|f>$ is the ground state without holes with all Cu spins ${\bf s}$
having the down projection. The state (\ref{6}) in the form (\ref{19})
is explicitly a one-particle state. One can easily prove that the state
(\ref{19}) is the eigenstate of the Hamiltonian (\ref{4}) in the form
(\ref{18}). For this we note that the operator $\hat{Z}_{l}({\bf k})$
can be represented in the form
\begin{eqnarray}
\label{X19}
\hat{Z}_{l}({\bf k})=\sqrt{2}(\alpha({\bf k})h^{+}_{l}+
\beta({\bf k})P^{+}_{l})\hat{s}_{l\downarrow} \ \ \ and \ \ \ H|f>=0.
\end{eqnarray}
Then commuting the  Hamiltonian (\ref{18}) with the operator
$\hat{Z}_{l}({\bf k})$ (\ref{X19}) and using the identity
\begin{eqnarray}
\label{Y19}
\hat{s}_{l}\hat{s}_{l'}|f>=(1/2)\hat{s}_{l}|f> \ \ \ for \ \ l \neq l'
\end{eqnarray}
we can obtain the expression (\ref{8}) for the coefficients $\alpha({\bf k}),
\beta({\bf k})$.

\section{Reduction to the generalized t~-~J model}
\clequ
\subsection{Transformation of the Hamiltonian to the Hubbard form} \

For deduction of the low-energy Hamiltonian it will be convenient to
transform the primary Hamiltonian (\ref{4}) to the form containing
exclusively the Hubbard operators. Such transformation is based on
a solution of the local or cluster problem for one electron or hole, when
cluster energy levels and cluster eigenfunctions are found.
After this we can express all operators contained in the primary
Hamiltonian (\ref{4}), such as $X^{22}_{l}, X^{\alpha2}_{l},X^{2\alpha}_{l},
P^{+}_{l\alpha}, P_{l\alpha}$, through the Hubbard operators
$X^{a'b'}_{l} = |a'><b'|$ characterizing cluster system, where $|a'>, |b'>$
are eigenfunctions of the cluster problem. Since only low-energy levels
of the cluster problem are essential for description of low-energy excitations
such transformation creates the basis for such description.

For realization of the program described above, let us introduce the Wannier
representation for the oxygen hole operators $P^{+}_{l\alpha},
P_{l\alpha}$ \cite{Zh1}
\begin{eqnarray}
\label{20}
&&P_{l}=\sum_{l'}\lambda(l,l')q_{l'}, \ \ \
P^{+}_{l}=\sum_{l'}\lambda(l,l')q^{+}_{l'}
\nonumber\\
&&\lambda(l,l')=\sum_{{\bf k}}\sqrt{(1+\gamma_{{\bf k}})}
\exp(i{\bf k}({\bf r}_{l}-{\bf r}_{l'})).
\end{eqnarray}
Since the Wannier-oxygen operators $q_{l}, q^{+}_{l}$ are independent at
different sites, the primary Hamiltonian (\ref{4}) can be expressed
through them. After this, the local or cluster problem can be solved. But
we will use another method of deduction of the extended Hubbard Hamiltonian.
This method is based on the use of the representation (\ref{17}) for Hubbard
operators. Hence we substitute the representation (\ref{20}) for
$P^{+}_{l\alpha}, P_{l\alpha}$ into the Hamiltonian (\ref{18}) and get
\begin{eqnarray}
\label{21}
&&H_{pd}=\epsilon_{p}\sum_{l}(q^{+}_{l}q_{l})+
\epsilon_{d}\sum_{l}h^{+}_{l}\hat{s}_{l}h_{l}+
\nonumber\\
&&2\sqrt{2}t\sum_{ll'}\lambda(l,l')(h^{+}_{l}\hat{s}_{l}q_{l'}+
q^{+}_{l'}\hat{s}_{l}h_{l}).
\end{eqnarray}
For solving the one-site problem we divide the operators
$q^{+}_{l\alpha}, q_{l\alpha}$ into the singlet and triplet parts
\begin{eqnarray}
\label{22}
&&q_{l}=q^{s}_{l}+q^{t}_{l},\ \ \ q^{s}_{l}=\hat{s}_{l}q_{l},\ \ \
q^{t}_{l}=\hat{t}_{l}q_{l},
\nonumber\\
&&\hat{s}_{l}=(1/4)(1-2{\bf s}_{l}{\bf \sigma}),\ \ \
\hat{t}_{l}=(1/4)(3+2{\bf s}_{l}{\bf \sigma}),\ \ \
\hat{t}_{l}+\hat{s}_{l}=1.
\end{eqnarray}
Then the one-site Hamiltonian has a simple quadratic form
\begin{eqnarray}
\label{23}
H^{0}_{pd}=\sum_{l}(\epsilon_{p}q^{+}_{l}\hat{s}_{l}q_{l}+
\epsilon_{d}h^{+}_{l}\hat{s}_{l}h_{l}+
\epsilon_{p}q^{+}_{l}\hat{t}_{l}q_{l}+
2\sqrt{2}t\lambda_{0}(h^{+}_{l}\hat{s}_{l}q_{l}+
q^{+}_{l}\hat{s}_{l}h_{l}))
\end{eqnarray}
and can be easily diagonalized
\begin{eqnarray}
\label{24}
H^{0}_{pd}=\sum_{l}(E_{-}c^{+}_{l}\hat{s}_{l}c_{l}+
E_{+}b^{+}_{l}\hat{s}_{l}b_{l}+
\epsilon_{p}q^{+}_{l}\hat{t}_{l}q_{l}),
\end{eqnarray}
where $E_{\pm}=\pm r$ with $r=\sqrt{\Delta^{2}+8\lambda^{2}_{0}
t^{2}} \ $ and $\lambda_{0} \equiv \lambda_{00}$.
New Fermi operators $c^{+}_{l}, c_{l}, b^{+}_{l}, b_{l}$ have the form
\begin{eqnarray}
\label{25}
c_{l}=bq_{l}-ah_{l},\ \ \ b_{l}=aq_{l}+bh_{l},
\end{eqnarray}
where $a=2\sqrt{2}t\lambda_{0}/\sqrt{2r(r+\Delta)},
\ \ b=\sqrt{(r+\Delta)/2r}$. The additional part of the
Hamiltonian $H^{int}_{pd}$ can be represented in the form
\begin{eqnarray}
\label{26}
&&H^{int}_{pd}=H^{1}_{pd}+H^{2}_{pd},\ \ \
H^{1}_{pd}=-4\sqrt{2}abt\sum_{ll'}\lambda_{ll'}(c^{+}_{l}\hat{s}_{l}
\hat{s}_{l'}c_{l'}-b^{+}_{l}\hat{s}_{l}\hat{s}_{l'}b_{l'})
\nonumber\\
&&H^{2}_{pd}=2\sqrt{2}t\sum_{ll'}\lambda_{ll'}((b^{2}-a^{2})
c^{+}_{l}\hat{s}_{l}\hat{s}_{l'}b_{l'}-
ac^{+}_{l}\hat{s}_{l}\hat{t}_{l'}q_{l'}+
bb^{+}_{l}\hat{s}_{l}\hat{t}_{l'}q_{l'}+h.c.).
\end{eqnarray}
Here and below we will separate $\lambda_{0}$ from $\lambda_{ll'}$
for $l \neq l'$ and suppose that all summations over $l, l'$ are
performed for $l \neq l'$. The Hamiltonians $H^{0}_{pd}$ and $H^{1}_{pd}$
sufficiently correctly describe the lower c-singlet band position and
c-singlet hopping to the nearest-neighbor sites. If we exclude the double
occupancy of the c-singlet sites, we can reduce this part of the
Hamiltonian $H^{0}_{pd}$ and $H^{1}_{pd}$ to the Hamiltonian of the t~-~J
model (\ref{1}) with the nearest-neighbor hopping. For this we
estimate the energy of the third additional hole on the elementary
Cu-O plaque. Choosing the hole wave function in the form
\begin{eqnarray}
\label{27}
|3l\alpha>=\xi_{3}P^{+}_{l\alpha}d^{+}_{l\uparrow}d^{+}_{l\downarrow}|0>+
\eta_{3}d^{+}_{l\alpha}P^{+}_{l\uparrow}P^{+}_{l\downarrow}|0>
\end{eqnarray}
and solving a simple variational problem we can get for the energy of
the three-hole state
\begin{eqnarray}
\label{28}
E_{3}=V/2+U_{p}/8-\sqrt{(\Delta+V/2-U_{p}/8)^{2}+4t^{2}\lambda_{0}^{2}}.
\end{eqnarray}
At $V=U_{p}=0$ the energy $E_{3}$ almost coincides with the top of the singlet
band on a ferromagnetic background, and the constants $V$ and $U_{p}$
give an additional gap.

Due to this estimation we can neglect the
contribution of the three-hole state in low-energy physics
and rewrite the Hamiltonian $H_{pd}$ in terms of the Hubbard operators.
For this we introduce Hubbard operators connected with the triplet states:
\begin{eqnarray}
\label{29}
&&X^{\pm 1\alpha}=(1/4)q^{+}[(1\pm 2s^{z})(1\pm \sigma^{z})+
s^{\pm}\sigma^{\pm}]_{\alpha}(1-\hat{n}^{q})
\nonumber\\
&&X^{0\alpha}=(1/2\sqrt{2})q^{+}[1-4s^{z}\sigma^{-}+
2{\bf s}{\bf \sigma}]\sigma^{x}_{\alpha}(1-\hat{n}^{q})
\nonumber\\
&&X^{\pm1 \pm 1}=(1/4)q^{+}(1-\hat{n}^{q})(1 \pm2s^{z})(1 \pm\sigma^{z})q
\nonumber\\
&&X^{00}=(1/4)q^{+}(1-\hat{n}^{q})[1-4s^{z}\sigma^{-}+ 2{\bf s}{\bf \sigma}]q.
\end{eqnarray}
Here the operators $X^{\mu \alpha}$ for $\alpha =\pm 1/2$ and $\mu =0,
\pm 1$ transform the spin-$\frac{1}{2}$ state with projection $\alpha$ into
the triplet spin-hole state with projection $\mu$ and $X^{\alpha \mu} =
(X^{\mu \alpha})^{+}$. The operators $X^{\mu \mu}$ act inside the
triplet states.

Substituting expressions of the operators $c^{+}_{l}\hat{s}_{l},
\hat{s}_{l}c_{l}, b^{+}_{l}\hat{s}_{l}, \hat{s}_{l}b_{l},
q^{+}_{l}\hat{t}_{l}, \hat{t}_{l}q_{l}$ in terms of the Hubbard operators,
after some calculations we get the following expression for the
Hamiltonian $H_{pd}$
\begin{eqnarray}
\label{30}
&&H^{0}_{pd}=\sum_{l}(E_{-}X^{cc}_{l}+E_{+}X^{bb}_{l})+
\epsilon_{p}\sum_{l\mu}X^{\mu \mu}_{l}
\nonumber\\
&&H^{1}_{pd}=-2\sqrt{2}abt\sum_{ll'}\lambda_{ll'}
(X^{c\alpha}_{l}X^{\alpha c}_{l'}-X^{b\alpha}_{l}X^{\alpha b}_{l'})
\nonumber\\
&&H^{2}_{pd}=\sqrt{2}t\sum_{ll'}\lambda_{ll'}[(b^{2}-a^{2})
X^{c\alpha}_{l}X^{\alpha b}_{l'}-
\nonumber\\
&&\sqrt{3}a(\alpha \beta \mu)X^{c\beta}_{l}X^{\alpha \mu}_{l'}+
\sqrt{3}b(\alpha \beta \mu)X^{b\beta}_{l}X^{\alpha \mu}_{l'}+ h.c.].
\end{eqnarray}
Here the Hubbard operators $X^{cc}_{l}, X^{bb}_{l}$,$X^{c\alpha}_{l}$,
$X^{\alpha c}_{l}$, $X^{b\alpha}_{l}$, $X^{\alpha b}_{l}$ are determined
by Eq.(\ref{17}) if we replace $h$-operators by $c$ and $b$-operators,
respectively. The operators $X^{\mu \mu}_{l}$, $X^{\mu \alpha}_{l}$,
$X^{\alpha \mu}_{l}$ are determined in Eq.(\ref{29}).
$(\alpha \beta \mu) \equiv <1/2\alpha|1/2 \beta, 1\mu >$ are Clebsh-Gordon
coefficients for angular momentum summation.

\subsection{Low-energy reduction of the extended Hubbard Hamiltonian} \

The Hamiltonian (\ref{30}) is equivalent to the primary Hamiltonian
(\ref{4}) but in this form it is substantially more convenient for
description of low-energy excitations. If we retain in Eq.(\ref{30}) only
the first two terms containing operators $X^{cc}_{l}, X^{c\alpha}_{l},
X^{\alpha c}_{l}$ and if we add the J-term from ({\ref{1}), we get
t~-~J the model with hopping to all sites. Indeed, the constants
$\lambda_{ll'}$ are different from zero for all sites and decrease rapidly
with increasing  distance between the sites $l$ and $l'$ :
\begin{eqnarray}
\label{31}
&&\lambda_{0} \equiv \lambda_{00}=0.9581,\ \ \ \ \ \
\lambda_{1} \equiv \lambda_{01}=0.1401,\ \ \ \
\lambda_{2} \equiv  \lambda_{11}=-0.02351,
\nonumber\\
&&\lambda_{3} \equiv \lambda_{02}=-0.01373,\ \
\lambda_{4} \equiv \lambda_{12}=0.006851,\ \
\lambda_{5} \equiv \lambda_{03}=0.003520,\ \
\nonumber\\
&&\lambda_{n,m} \Longrightarrow (-1)^{m+n+1}/
2\pi (n^{2}+m^{2}+nm)^{3/2} \ \ \ for \ \ n+m \gg 1,
\end{eqnarray}
where $\lambda_{nm} \equiv \lambda_{0,{\bf l}}$ for ${\bf l} =
n{\bf e}_{x}+m{\bf e}_{y}$. Since the constants $\lambda_{nm}$
decrease sufficiently rapidly with increasing $m+n$, we can construct the
perturbation theory over the Hamiltonian $H^{2}_{pd}$ with the help
of the Schrieffer-Wolff transformation
\begin{eqnarray}
\label{32}
H_{pd} \Rightarrow \tilde{H}_{pd}=\exp{(-S)}H_{pd}\exp{(S)}, \ \ \ S^{+}=-S.
\end{eqnarray}
In the first order of perturbation theory over $H^{2}_{pd}$ the generator
of transformation and the second-order correction to the Hamiltonian are
\begin{eqnarray}
\label{33}
[H^{0}_{pd},S]=-H^{2}_{pd}, \ \ \ \delta H_{pd}=(1/2)[H^{2}_{pd},S].
\end{eqnarray}
For the Hamiltonian $H_{pd} $ in the form (\ref{30}) the generator S
can be easily found using the properties of the Hubbard algebra
\begin{eqnarray}
\label{34}
S=\sqrt{2}t\sum_{ll'}\lambda_{ll'}[\frac{b^{2}-a^{2}}
{E_{+}-E_{-}}X^{c\alpha}_{l}
X^{\alpha b}_{l'}-\frac{\sqrt{3}a}{\epsilon -E_{-}}
(\alpha \beta \mu)X^{c \beta}_{l}X^{\alpha \mu}_{l'} - h.c.].
\end{eqnarray}
We retain here the contribution to the generator S essential for the
correction to the  lower c-singlet Hamiltonian. On the basis of
this formula for the generator S, using an explicit form of the
parameters $E{\pm}, a, b$ and the summation formulae for the Clebsh-Gordon
coefficients, one can get the expression for the correction
$\delta H_{pd} $ to the Hamiltonians $H^{0}_{pd}$ and $H^{1}_{pd}$
\begin{eqnarray}
\label{35}
&&\delta H_{pd}=(t^{2}/r)\sum_{lnl'}\lambda_{ln}\lambda_{nl'}
[(1-\Delta^{2}/r^{2})X^{c\alpha}_{l}X^{\alpha \beta}_{n}X^{\beta c}_{l'}
\nonumber\\
&&-2\hat{N}_{n}X^{c\alpha}_{l}X^{\alpha c}_{l'}],
\ \ \ \hat{N}_{n}\equiv X^{\uparrow\uparrow}_{n}+
X^{\downarrow\downarrow}_{n}
\end{eqnarray}
The presence of the operators $X^{\alpha \beta}_{n}$ in  Eq.(\ref{35})
reflects the Fermi statistics of holes. Hopping from  the site $l$ to
the site $l'$ through the site $n$ depends on the filling and the
spin state of
a hole at the site $n$.

At this final step we can add the Hamiltonian $\delta H_{pd}$ to
the Hamiltonians $H^{0}_{pd}, H^{1}_{pd}$ and obtain the Hamiltonian
correctly describing the energy of the lower c-singlet and its hopping to the
nearest-neighbors. Using the identity
\begin{eqnarray}
\label{36}
X^{\alpha \beta}_{n}=(\hat{N}_{n}/2)\delta_{\alpha \beta}+
{\bf s}_{n}{\bf \sigma}_{\beta \alpha},
\end{eqnarray}
where ${\bf s}_{n}$ is the Cu spin-$\frac{1}{2}$ operator, one can get
\begin{eqnarray}
\label{37}
&&H^{01}_{pd}=E_{0}\sum_{l}X^{cc}_{l}+
t_{1}\sum_{<ll'>}X^{\alpha c}_{l}X^{c\alpha}_{l'}+
E_{0N}\sum_{ln}\lambda_{ln}^{2}\hat{N}_{n} X^{cc}_{l}+
\nonumber\\
&&\sum_{<ll'>n}\lambda_{ln}\lambda_{nl'}[t_{1N}
\hat{N}_{n}X^{\alpha c}_{l}X^{c \alpha}_{l'}+
t_{1S}{\bf s}_{n}X^{\alpha c}_{l}{\bf \sigma}_{ \alpha \beta}
X^{c \beta}_{l'}],
\end{eqnarray}
where $E_{0}=-r, \ t_{1}=4\lambda_{0}\lambda_{1}t^{2}/r$,
 $E_{0N}=-(3+\Delta^{2}/r^{2})(t^{2}/r)$, $t_{1N}=(t^{2}/2r)
(3+\Delta^{2}/r^{2})$, $t_{1S}=-(t^{2}/r)(1-\Delta^{2}/r^{2})$  .
Using the representation (\ref{tjt}) of the Hubbard  operators in terms of
the primary electron operators $c_{l\alpha}^{+},c_{l\alpha}$, we can
rewrite the expression for the Hamiltonian (\ref{37}) in a more usual form
\begin{eqnarray}
\label{37a}
&&H^{01}_{pd}=E_{0}\sum_{l}(1-\hat{n}_{l}^{c}+\hat{d}_{l}^{c})+
t_{1}\sum_{<ll'>}(\tilde{c}^{+}_{l}\cdot \tilde{c}_{l'})+
\nonumber\\
&&E_{0N}\sum_{ln}\lambda_{ln}^{2}(\hat{n}_{n}^{c}-2\hat{d}_{n}^{c})
(1-\hat{n}_{l}^{c}+\hat{d}_{l}^{c})+
\sum_{<ll'>n}\lambda_{ln}\lambda_{nl'}[t_{1N}
(\hat{n}_{n}^{c}-2\hat{d}_{n}^{c})(\tilde{c}^{+}_{l}\cdot \tilde{c}_{l'})+
\nonumber\\
&&t_{1S}(\tilde{c}^{+}_{n}{\bf \sigma}\tilde{c}_{n})
(\tilde{c}^{+}_{l}{\bf \sigma}\tilde{c}_{l'})],\ \  \
\hat{d}_{l}^{c}\equiv
 c^{+}_{\uparrow}c_{\uparrow}c^{+}_{\downarrow}c_{\downarrow}.
\end{eqnarray}

The first two
terms of these Hamiltonians (\ref{37}),(\ref{37a})
 coincide with the first two terms of the t~-~J
Hamiltonian (\ref{1}). The second two terms represent the second-order
corrections which depend on the filling and the spin state of the
neighbor sites.
The relative magnitude of these additional terms is
approximately 10\% of the first two terms. In this case
summation over index $n$ can be limited by the nearest-neighbors,
next-nearest-neighbors and next-to-next-nearest neighbors of the sites $l$
and $l'$. A more detailed comparison of the relative contribution of
different terms in the Hamiltonian (\ref{37}) for the case of a
ferromagnetic background is presented below.
If we add to the Hamiltonian (\ref{37}) the J-term with the
nearest-neighbors exchange:
$ J=4t^{4}(U_{d}-2\Delta )^{-2}(1/U_{d}+1/(U_{d}-2\Delta))
\simeq 0.13$eV (see \cite{Em1,Zh1}), we get the effective
one-band Hamiltonian for our case.

\subsection{The structure of hopping to the next neighbors} \

Some works \cite{En1,Ba1,Da1,Ch1,Ho1} on the t~-~J model consider
different generalization of the usual  t~-~J Hamiltonian. The reason for
such consideration is a dependence of the one-particle energy for
the antiferromagnetic spin ordering on the details of the Hamiltonian.
In some works the exchange Hamiltonian for the next neighbors (frustration)
was considered \cite{Ba1}. Such terms were deduced from the three-band
Hamiltonian in Ref. \cite{Yu2}. The last two terms in the Hamiltonian
(\ref{37}) represent the corrections to the energy-level position
and to the hopping to the nearest-neighbors. But the three-band Hamiltonian
also generates hopping to the next-nearest-neighbors. The
structure of these
additional terms in the total Hamiltonian is following
\begin{eqnarray}
\label{38}
H^{23}_{pd}=\sum_{i=2,3}[t_{i}\sum_{<ll'i>}X^{\alpha c}_{l}
X^{c \alpha}_{l'}+\sum_{<ll'i>n}\lambda_{ln}\lambda_{nl'}
(t_{iN}N_{n} X^{\alpha c}_{l}X^{c \alpha}_{l'}+
t_{iS}{\bf s}_{n}X^{\alpha c}_{l}{\bf \sigma}_{\alpha \beta}
X^{c \beta}_{l'})],
\end{eqnarray}
where $t_{i}=4t^{2}\lambda_{i}\lambda_{0}/r$,
  $t_{iN}=(t^{2}/r)(3+\Delta^{2}/r^{2})$,
 $t_{iS}=-2(t^{2}/r)(1-\Delta^{2}/r^{2})$  for $i=2,3$ and $<ll'i>$
denotes summation over the second or third neighbors. The last term
formally coincides with the last term of the Hamiltonian $H_{01}$
but summation over $l$ and $l'$ is performed over the second and third
neighbors. The summation over $n$ is performed  over the sites
nearest to $l$ and $l'$ . The physical interpretation
of the Hamiltonian is similar
to that of $H_{01}$. Corrections of the third- and the fourth-order to the
Hamiltonian $H_{01}+H_{23}$ are considered in Appendix.

\section{Comparison with the exact solution on a ferromagnetic background}
\clequ
\subsection{Comparison of the second-order Hamiltonian} \

For one-hole problem on a ferromagnetic background, the Hamiltonian (\ref{37})
is substantially simplified and can be represented in the form
\begin{equation}
\label{39}
H_{01}=E^{f}_{0}\sum_{l}X^{cc}_{l}+t^{f}_{1}\sum_{<ll'>}X^{\downarrow c}_{l}
X^{c \downarrow}_{l'},
\end{equation}
where the parameters $E^{f}_{0}, t^{f}_{1}$ have the form
\begin{eqnarray}
\label{40}
&&E^{f}_{0}=-r -(4t^{2}/r)(3+\Delta^{2}/r^{2})(\lambda^{2}_{1}+
\lambda^{2}_{2}+\lambda^{2}_{3})
\nonumber\\
&&t^{f}_{1}=(2\lambda_{1}t^{2}/r)(2\lambda_{0}+(1+\Delta^{2}/r^{2})
(2\lambda_{2}+\lambda_{3})).
\end{eqnarray}
We want to discuss two questions: (1) the relative magnitude of the
second-order corrections and (2) a comparison with the exact result.
We can compare these parameters $E^{f}_{0}, t^{f}_{1}$ of the approximate
Hamiltonian (\ref{37}) with the exact parameters $E^{ex}_{0}, t^{ex}_{1}$
of the exact solution on the ferromagnetic background
\begin{eqnarray}
\label{41}
t^{ex}_{i}=-\sum_{\bf{k}}\epsilon _{{\bf k}}\cos{({\bf k}{\bf r}_{i})},\ \ \
E^{ex}_{0} =-t^{ex}_{0},
\end{eqnarray}
where the energy of a hole on a ferromagnetic background
$\epsilon _{{\bf k}}$ is represented by a very simple Eq.(\ref{7}),
and ${\bf r}_{i}$ is the vector from the origin to the i-neighbor.
At the first step let us compare Eq.(\ref{7}) and Eq.(\ref{39})
in two limiting cases $\Delta \gg t$ and $\Delta \ll t$. In the first
case $\Delta \gg t$ we have the exact parameters $E^{ex}_{0}, t^{ex}_{1}$
\begin{eqnarray}
\label{42}
E^{ex}_{0}=-\Delta-4t^{2}/\Delta,\ \ \
t^{ex}_{1}=t^{2}/2\Delta
\end{eqnarray}
and for the approximate parameters $E^{f}_{0}, t^{f}_{1}$
\begin{eqnarray}
\label{43}
&&E^{f}_{0}=-\Delta-(4t^{2}/\Delta)(\lambda^{2}_{0}+
4(\lambda^{2}_{1}+\lambda^{2}_{2}+\lambda^{2}_{3}))
\simeq -\Delta- 3.998t^{2}/\Delta,
\nonumber\\
&&t^{f}_{1}=(4\lambda_{1}t^{2}/2\Delta)(\lambda_{0}+
2\lambda_{2}+\lambda_{3}) \simeq 0.5031t^{2}/\Delta.
\end{eqnarray}
We can see the agreement up to the third digit.
The relative magnitudes of the corrections to $E^{f}_{0}$ and $t^{f}_{1}$
 are  0.089 and 0.061 respectively.
 In the opposite case
$t \gg \Delta $ we can compute the integrals (\ref{41}) for
$E^{ex}_{0}, t^{ex}_{1}$
\begin{eqnarray}
\label{44}
E^{ex}_{-}=-2.8053t, \ \ \ \ t^{ex}_{1}=0.1801t,
\end{eqnarray}
and have for the approximate case
\begin{eqnarray}
\label{45}
&&E^{f}_{0}=-2\sqrt{2}\lambda_{0}t(1+3(\lambda^{2}_{1}+\lambda^{2}_{2}+
\lambda^{2}_{3})/\lambda^{2}_{0})=-2.8001t
\nonumber\\
&&t^{f}_{1}=2\sqrt{2}\lambda_{1}(1+(2\lambda_{2}+\lambda_{3})/\lambda_{0})=
0.19118t.
\end{eqnarray}
The agreement between $E^{ex}_{0}$ and $E^{f}_{0}$ is also
up to the third digit, but agreement between $t^{ex}_{1}$
and $t^{f}_{1}$ is of about 5\%.
The relative magnitudes of the corrections in this case to
 $E^{f}_{0}$ and $t^{f}_{1}$  are  0.066 and 0.031 respectively.
 In Table 1 we give the
values of the parameters $E^{ex}_{0}, E^{f}_{0},
t^{ex}_{1}$, and $t^{f}_{1}$ for different values of the
$\Delta^{2}/t^{2}$ ratio.

TABLE 1.{\small Parameters of the effective t~-~J model
on a ferromagnetic
background for the exact solution $E^{ex}_{0}, t^{ex}_{1}$ and
for the reduced Hamiltonian (\ref{39}) $E^{f}_{0}, t^{f}_{1}$ as a function
of $\Delta^{2}/t^{2}$.}
\vskip 0.5 cm
\begin{center}
\begin{tabular}{||c||c|c|c|c|c||} \hline
$\Delta^{2}/t^{2}$ & 0.01 & 0.1 & 1.0 & 10.0 & 100.0 \\ \hline
$|E^{ex}_{0}|$  & 2.8077 & 2.8233 & 2.9808 & 4.2360 & 10.3919 \\ \hline
$|E^{f}_{0}|$   & 2.8758 & 2.8917 & 3.0460 & 4.2835 & 10.4117 \\ \hline
$t^{ex}_{1}$  & 0.1800 & 0.1789 & 0.1691 & 0.1182 & 0.04807 \\ \hline
$t^{f}_{1}$   & 0.1865 & 0.1854 & 0.1751 & 0.1211 & 0.04858 \\ \hline
\end{tabular}
\end{center}
\vskip 0.5 cm

\subsection{More detailed comparison of the fourth-order Hamiltonian} \

We will make a more detailed comparison of the parameters of the effective
Hamiltonian on a ferromagnetic background with the exact solution in the
practically important limit  $\Delta \ll t$. The corrections of the
third order to the one-band Hamiltonian, obtained in Appendix,
lead to the following
corrections to the parameters $E^{f}_{0}, t^{f}_{1}$ of the Hamiltonian
(\ref{39})
\begin{eqnarray}
\label{46}
&&\delta E^{f}_{0}=(\lambda_{2}t/\sqrt{2}\lambda^{2}_{0})
(3\lambda^{2}_{1}+2\lambda_{2}\lambda_{3})-
54t\lambda^{4}_{1}/16\sqrt{2}\lambda^{3}_{0} \simeq -0.0021t,
\nonumber\\
&&\delta t^{f}_{1}=-(\lambda_{1}t/4\sqrt{2}\lambda^{2}_{0})
(17\lambda^{2}_{1}+20\lambda^{2}_{2}+18\lambda^{2}_{3}) \simeq -0.00939t.
\end{eqnarray}
Summing up these expressions for $\delta E^{f}_{0}$ and $\delta t^{f}_{1}$
with $E^{f}_{0}$ and $t^{f}_{1}$ from Eq.(\ref{45}) we get
\begin{eqnarray}
\label{47}
E^{f}_{0}=-2.8023t,\ \ \ \ t^{f}_{1}=0.1824t.
\end{eqnarray}
Comparing with the exact values $E^{ex}_{0}, t^{ex}_{1}$ of Eq.(\ref{44})
we see an excellent quantitative agreement.

We also will compare the hopping Hamiltonian for the second and the third
neighbors on a ferromagnetic background
\begin{equation}
\label{48}
H_{23}=\sum_{i=2,3}t^{f}_{i}\sum_{<ll'i>}X^{\downarrow c}_{l}
X^{c \downarrow}_{l'}
\end{equation}
with the exact solution (\ref{6}). The exact hopping parameters
$t^{ex}_{2}, t^{ex}_{3}$ on a ferromagnetic background are equal to
\begin{eqnarray}
\label{49}
t^{ex}_{2}=-0.01177t, \ \ \ \ t^{ex}_{3}=-0.00603t.
\end{eqnarray}
We represent the expression for the approximate values of the parameters
$t^{f}_{2}, t^{f}_{3}$ in the form
\begin{eqnarray}
\label{50}
t^{f}_{i}=t^{(1)}_{i}+t^{(2)}_{i}+t^{(3)}_{i}+t^{(4)}_{i}
\end{eqnarray}
where $t^{(j)}_{i}$ for j=1,2,3,4 correspond to the contribution of
the j-th order of the perturbation theory. The explicit expressions for
the parameters $t^{(j)}_{2}/t$, as is shown in Appendix, are following
\begin{eqnarray}
\label{51}
&&t^{(1)}_{2}/t=\sqrt{2}\lambda_{2}=-0.0332,
\nonumber\\
&&t^{(2)}_{2}/t=(1/\sqrt{2}\lambda_{0})(\lambda^{2}_{1}+
2\lambda_{2}\lambda_{3}+2\lambda_{1}\lambda_{4})=0.0164,
\nonumber\\
&&t^{(3)}_{2}/t=-(1/4\sqrt{2}\lambda^{2}_{0})(26\lambda^{2}_{1}\lambda_{2}+
8\lambda^{2}_{1}\lambda_{3}+24\lambda_{2}\lambda^{2}_{3}+
17\lambda^{3}_{2})=0.0028,
\nonumber\\
&&t^{(4)}_{2}/t=13\lambda^{4}_{1}/4\sqrt{2}\lambda^{3}_{0}=0.0010.
\end{eqnarray}
and in a more compact form for $t^{(j)}_{3}/t$
\begin{eqnarray}
\label{52}
t^{(j)}_{3}/t=(-0.0194, 0.0095, 0.0021, 0.0006)
\end{eqnarray}
As a result we have
\begin{eqnarray}
\label{53}
t^{f}_{2}=-0.013t, \ \ \ \ \ t^{f}_{3}=-0.0072t.
\end{eqnarray}
In this case we can see that the agreement between the exact  and
approximate values is of the order of 10\%.
We  have an substantial  compensation of the  direct hopping  constants
$t^{(1)}_{2}, t^{(1)}_{3}$ up to the final values $t^{f}_{2}, t^{f}_{3}$
on a ferromagnetic background due to the higher-order  correction.   This
means that the corrections to hopping on the  next  neighbors
have a very complicated nature  and include hopping processes
depending on the Cu spin states at the neighbor sites.

\section{ Conclusion}
\clequ \

It has been shown that in the case of the three-band model
for $\epsilon \gg U_{d}
-\epsilon$ the  Bloch waves constructed from the local Cu-O singlets
are the ground states for one-hole excitations.  The Zhang-Rice Cu-O
singlets  form the basis for reduction of the three-band model to the
single-band generalized
t~-~J model.  The method of the reduction  developed in  this  work  is
rather specific and is based on the representations  of  the  Hubbard
operators in terms of the Fermi and spin-$\frac{1}{2}$ operators.
This reflects the history of  work on the  paper.

In  fact,  the
method of  obtaining  a  single-band  Hamiltonian  is  sufficiently
general: at the first step the  claster  problem  is  solved  and  local
electron (hole) energy levels and wave functions are found
 with  correlations being taken into account .
  At the second  step   the   initial
Hamiltonian can be expressed in terms of the Hubbard  operators which
transfer these states in each other, including the ground  state.
This Hamiltonian includes hopping terms which  describe
a hole transition from one lattice site to  another, including mixing
of the local energy positions.
     If such mixing is small, at the third  step  with  the  help  of
the Schrieffer-Wolff  transformation    one   can   get   the  single-band
Hamiltonian for description of the low-energy excitations.

     In the framework of such approach one can consider the general  case  of
the three-band model parameters when $\epsilon, U_{d}-\epsilon$
and $4\sqrt{2}t$ are of the   same   order   of    magnitude.
One can also take the direct oxygen-oxygen hopping into consideration .

     We want to stress that the singlet structure  of  hole  excitations
based on the Wannier functions
provides a very low energy of hole excitations.  In the case
considered by Lovtzov and Yushankhai \cite{Yu1} and in the case
discussed in this work the situation is similar:  the position  of
the bottom     of the  hole  singlet  band, measured from the  middle
of the spacing between the oxygen and copper local levels, is
 equal to (\ref{37})
\begin{eqnarray}
\label{54}
E_{b}=-r-\xi t_{1} \simeq -2\sqrt{2}(1+0.067\xi)t \simeq -3.1t,
\end{eqnarray}
where $\xi$ is the parameter of the order of a unity which
describes the  dependence
of the band-bottom position on the type of  magnetic
ordering of the copper spins.  For the antiferromagnetic ordering
$\xi =3.13-2.83(J/t)^{0.73}$ \cite{Da2} and for our case J=0.13eV and
$t_{1}=0.27$eV $\rightarrow \ \xi=1.47$ and $E_{b}=-3.1t$. It is necessary
to stress that this value of the bottom of the band position is very low.
The  attempts to develop the physics of three-band model in  terms  of
the slave-boson approach \cite{Ca1,Gr1,Gr2,Fe1} give
$E_{b}=-2\sqrt{2}r_{0}t$ with $r_{0} < 0.6$ which yields $E_{b}\ =\ -1.7t$
for the bottom of band for some variants of the spin-liquid  state.
This state is positioned sufficiently  high. It  is  unlikely  that  the
gain in the exchange energy of the  spin-liquid  state  due  to  the
presence of holes makes the energy of such type of state  lower  than
the energy of the singlet band.

Actually the  gain in  the  exchange
energy has scale J while the band position has scale $t$, but $t/J \simeq 10$.
Of course we can not prove a theorem that the hole singlet band has
the lowest possible energy for the three-band  model  in  the  actual
region of parameters.  However, in our case such theorem has been  proved
for a ferromagnetic background, and we believe that the consideration of the
general magnetic state does not change the situation.

In our case ($U_{d}-\epsilon \ll \epsilon$) the fundamental parameter of the
t~-~J model $J/t_{1} \simeq J/0.19t \simeq 0.45$. This estimation of the
ratio $J/t_{1}$ correlates fairly well with another estimations of
this  ratio \cite{Fl1,Ri1,Hy1}.

Hopping to the next neighbors has complicated nature depending  on
spin  states  of  the neighbor   copper   ions   and   can   not   be
expressed by a simple $t'$-term with hopping to the next-nearest-neighbors.
The  order  of magnitude of these terms in  the  Hamiltonian  is $(0.02
\div 0.03)t$ which is 10\% of $t_{1}$.

\section{Acknowledgments} \

We would like to thank O.P.Sushkov for stimulating  discussions
and V.Yu.Yshankhai for useful conversations.
This work was supported partly by the Counsel on Superconductivity
of Russian Academy of Sciences, Grant No. 90214.

\section{Appendix A}
\clequ \

We derive here the third- and fourth-order corrections to the effective
Hamiltonian (\ref{30}), and the forms of the corresponding energy additions.
We restrict ourselves by the case $\Delta=0$ and  a ferromagnetic background.
The full Hamiltonian in terms of the Hubbard operators (\ref{30})
can be expressed in more convenient terms. Let us introduce:
\begin{eqnarray}
\label{a1}
&&D=\sum_{ll'}\lambda_{ll'}(X^{c\alpha}_{l}X^{\alpha c}_{l'}-
X^{b\alpha}_{l}X^{\alpha b}_{l'}),
\nonumber\\
&&F=\sum_{ll'}\lambda_{ll'}(\alpha\beta\mu)(X^{c\beta}_{l}
X^{\alpha\mu}_{l'}-X^{\mu\alpha}_{l}X^{\beta b}_{l'}),
\nonumber\\
&&G=\sum_{ll'}\lambda_{ll'}(\alpha\beta\mu)(X^{\mu\alpha}_{l}
X^{\beta c}_{l'}-X^{b \beta}_{l}X^{\alpha\mu}_{l'}).
\end{eqnarray}
One can check that
\begin{eqnarray}
\label{a37}
[H_{0},D]=0, \ \ \ [H_{0},F]=-\eta tF, \ \ \ [H_{0},G]=\eta tG,
\end{eqnarray}
where $\eta=2\sqrt{2}\lambda_{0}$ . Hence, G can be named as 'raising' and F
'lowering' operators, because G transfers low-singlet state to triplet one,
triplet to high-singlet, while F acts in the opposite way.

In these (\ref{a1}) terms the Hamiltonian (\ref{30}) has the form:
\begin{eqnarray}
\label{a2}
&&H_{1}=-\sqrt{2}tD,\ \ \  H_{2}=-\sqrt{3}t(F+G).
\end{eqnarray}
The first-order generator of the Schrieffer-Wolff transformation
and the second-order
 term of the effective Hamiltonian are given by:
\begin{eqnarray}
\label{a3}
&&S_{1}=-(\sqrt{3}/\eta)(F-G), \ \
\delta H^{(2)}=-(3t/\eta)(FG-GF).
\end{eqnarray}
  By projecting out highly excited states, the second-order term in
$\delta H^{(2)}$ can be obtained. Equation for the second-  and third-order
generators of the Schrieffer-Wolff transformation and for the third- and
fourth-order corrections to the  interaction are \cite{Bi1} :
\begin{eqnarray}
\label{a4}
&&[H_{0},S_{2}]=-[H_{1},S_{1}], \ \ \ \
[H_{0},S_{3}]=-[H_{1},S_{2}]-(1/3)[[H_{2},S_{1}],S_{1}],
\nonumber\\
&&\delta H^{(3)}=(1/2)[H_{2},S_{2}], \ \ \ \
\delta H^{(4)}=(1/2)[H_{2},S_{3}]-(1/24)[[[H_{2},S_{1}],S_{1}],S_{1}]
\end{eqnarray}
in terms of D,F, and G we get :
\begin{eqnarray}
\label{a5}
&&S_{2}=-(\sqrt{6}/\eta^{2})[D(F+G)-(F+G)D],
\nonumber\\
&&S_{3}=-(2\sqrt{3}/\eta^{3})[(G-F)D^{2}-2D(G-F)D+
\nonumber\\
&&D^{2}(G-F)-2GFG+2FGF-GFF+FGG+GGF-FFG]
\end{eqnarray}
so that :
\begin{eqnarray}
\label{a6}
&&\delta H^{(3)}=(3t/\sqrt{2}\eta^{2})[D(F+G)^{2}+(F+G)^{2}D-
2(F+G)D(F+G)].
\end{eqnarray}
Since we are interested in the low-energy states, all terms with  F to the
right
and G to the left can be omitted. Also the third term in (\ref{a6})
can be removed
because the triplet state does not hope. Thus we get the effective
\begin{eqnarray}
\label{a7}
&&\delta H^{(3)}=(3t/\sqrt{2}\eta^{2})[DFG+FGD].
\end{eqnarray}
Corresponding corrections to the effective hopping and the energy on
a ferromagnetic background have the form:
\begin{eqnarray}
\label{a8}
&&\delta t^{(3)f}=(\sqrt{2}t/\eta^{2})[\sum_{ll'}\lambda_{il} \
\lambda_{ll'} \ \lambda_{lj}+2\lambda_{ij} \ \sum_{l}(\lambda_{il}^{2}+
\lambda_{lj}^{2})+3\lambda_{ij}^{3}]
\nonumber\\
&&\delta E^{(3)f}_{0}=(\sqrt{2}t/\eta^{2})\sum_{ll'}\lambda_{il}
\ \lambda_{ll'} \ \lambda_{l'i}.
\end{eqnarray}
  By substitution (\ref{a5}) in (\ref{a4}) and keeping low-energy terms,
we have
\begin{eqnarray}
\label{a9}
&&\delta H^{(4)}=(3t/2\eta^{3})[6FGFG-3F^{2}G^{2}-FGD^{2}-D^{2}FG].
\end{eqnarray}
The fourth-order corrections to the hopping parameter and energy
will be
\begin{eqnarray}
\label{a10}
&&\delta t^{(4)f}=-(t/2\eta^{3})[3\sum_{lnm}\lambda_{il} \
\lambda_{ln} \ \lambda_{nm} \ \lambda_{mj}+5\sum_{ln}\lambda_{il} \
\lambda_{lj}(\lambda_{in}^{2}+\lambda_{jn}^{2})+
\nonumber\\
&&3\sum_{ln}\lambda_{il} \ \lambda_{lj} \ \lambda_{ln}^{2}+
5\sum_{l}(\lambda_{il}^{3} \ \lambda_{lj}+\lambda_{il} \ \lambda_{jl}^{3})+
\nonumber\\
&&3\lambda_{ij} \ \sum_{ln}(\lambda_{il} \ \lambda_{ln} \ \lambda_{ni}+
\lambda_{jl} \ \lambda_{ln} \ \lambda_{jn})+13\lambda_{ij}^{2} \
\sum_{l}\lambda_{il} \ \lambda_{lj}]
\nonumber\\
&&\delta E^{(4)f}_{0}=-(3t/2\eta^{3})[\sum_{lnm}\lambda_{il} \
\lambda_{ln} \ \lambda_{nm} \ \lambda_{mi}+\sum_{ln}(\lambda_{il}^{2} \
\lambda_{ln}^{2}+\lambda_{il}^{2} \ \lambda_{in}^{2})+
\sum_{l}\lambda_{il}^{4}].
\end{eqnarray}
  For hopping to the first, second, and third neighbors corrections
may be easily calculated.

\end{document}